\documentstyle[epsf]{article}
\newcommand{\gsim}{ \raisebox{-.5ex}{\mbox{$\,\stackrel{>}{\sim}\,$}} }
\newcommand{\lsim}{ \raisebox{-.5ex}{\mbox{$\,\stackrel{<}{\sim}$\,}} }
\begin{document}
\begin{center}
{\large\bf Big Entropy Fluctuations in Nonequilibrium Steady State:\newline
                         A Simple Model with Gauss Heat Bath}\\[2mm]

                   Boris Chirikov\\
{\it Budker Institute of Nuclear Physics \\
        630090 Novosibirsk, Russia}\\[1mm]
        chirikov @ inp.nsk.su\\[5mm]
\end{center} 
\baselineskip=10pt

\vspace{2mm}

\begin{abstract} 
Large entropy fluctuations in a nonequilibrium steady state 
of classical mechanics
were studied
in extensive numerical experiments on a simple 2--freedom model 
with the so--called
Gauss time--reversible thermostat.
The local fluctuations (on a set of fixed trajectory segments) 
from the average heat entropy absorbed in thermostat were
found to be non--Gaussian. Approximately, the fluctuations can be discribed 
by a two--Gaussian
distribution with a crossover independent of the segment length
and the number of trajectories ('particles').
The distribution itself does depend on both, approaching the
single standard Gaussian distribution as any of those parameters increases.
The global time--dependent fluctuations 
turned out to be qualitatively different in that they have a
strict upper bound much less than the average entropy production.
Thus, unlike the equilibrium steady state, the recovery of the initial
low entropy becomes impossible, after a sufficiently long time, even in the
largest fluctuations. 
However, preliminary numerical experiments and the theoretical estimates
in the special case of the critical dynamics with superdiffusion
suggest the existence of infinitely many Poincar\'e 
recurrences to the initial state and beyond. This is a new interesting phenomenon to be farther
studied together with
some other open questions.
Relation of this particular example of nonequilibrium 
steady state to
a long--standing persistent controversy over statistical 'irreversibility',
or the notorious 'time arrow', is also discussed. In conclusion, 
an unsolved problem of the origin of the causality
'principle' is touched upon.
\end{abstract}

\vspace{1cm}

\section{Introduction:\newline
 Equilibrium vs. nonequilibrium steady state}
The fluctuations are inseparable part of the statistical laws.
This is well known since Boltzmann.
What is apparently less known are the peculiar properties of rare 
big fluctuations
(BF) as different from, and even opposite in a sense to, 
those of small stationary 
fluctuations.
Particularly, the former may be perfectly regular, on the average, 
symmetric in time with respect
to the fluctuation maximum, and described by simple kinetic
equations rather than by a sheer probability
of irregular 'noise'. Even though BF
are very rare they may be important
in many various applications (see, e.g., \cite{1} and references therein).
Besides, the correct understanding and interpretation of properties and 
the origin of BF may help (at last!) to settle
a strange long--standing persistent controversy over statistical 
'irreversibility' and the notorious 'time arrow'.

In the BF problem one should distinguish at least two 
qualitatively different classes of the fundamental 
(Hamiltonian, nondissipative) 
dynamical systems: those {\it with} and
{\it without} the statistical equilibrium, or {\it equilibrium steady state}
(ES).

In the former (simpler) case a BF consists of the two symmetric parts: 
the rise
of a fluctuation followed by its return, or relaxation,
back to ES (see Fig.1 below).
Both parts are described by the same kinetic (e.g., diffusion) equation,
the only difference being in the sign of time. This relates the
time--symmetric dynamical equations to the time--antisymmetric
{\it kinetic} (but not statistical!) equations. The principal difference
between the both, some times overlooked, is in that the kinetic equations
are widely understood as describing the relaxation only, i.e. 
{\it increase} of the entropy in a closed system whereas, in fact, they do so
for the rise of BF as well, i.e. for the entropy {\it decrease}.
All this was qualitatively known already to Boltzmann \cite{2}.
The first simple example of a symmetric BF
was considered by Schr\"odinger \cite{3}. Rigorous mathematical theorem
for the diffusive (slow) kinetics was proved by Kolmogorov in 1937
in the paper entitled 'Zur Umkehrbarkeit der statistischen Naturgesetze'
('Concerning reversibility of statistical laws in nature')
\cite{4} (see also
\cite{5}). Regrettably, the principal Kolmogorov theorem still remains
unknown to both the participants of hot debates over 'irreversibility' 
(see, e.g., 'Round Table on Irreversibility' in \cite{6})
as well as the physicists actually studying
such BF \cite{1}.

By now, there exists the well developed ergodic theory of dynamical
systems (see, e.g., \cite{7}). Particularly, it proves that the relaxation
(correlation decay, or mixing) proceeds eventually in both directions of time
for almost any initial conditions of a chaotic dynamical system.
However, the relaxation does not need to be always monotonic
which simply means
a BF on the way, depending on the initial conditions.
To get rid of such an apparently confusing (to many) 'freedom' one can take
a different approach to the problem: to start at arbitrary initial conditions
(most likely corresponding to ES), and see the BF dynamics and statistics.

At this point, it is essential to remind that the systems with ES allow for
very simple models in both the theoretical analysis as well as numerical
experiments which the latter are even more important.
In the present paper I will use one of the most simple and popular model 
specified
by the so--called Arnold cat map (see \cite{8,9}):
$$ 
   \begin{array}{ll}
   \overline{p}\,=\,p\,+\,x & mod\ 1 \\
   \overline{x}\,=\,x\,+\,\overline{p} & mod\ 1 
   \end{array} 
   \eqno (1.1)
$$
which is a linear canonical map on a unit torus. It has no parameters, and is
chaotic and even ergodic. The rate of the local exponential instability,
the Lyapunov exponent $\lambda =\ln{(3/2+\sqrt{5}/2)}=0.96$, implies
a fast (ballistic) kinetics with relaxation time $t_r\sim 1/\lambda\approx 1$.

A minor modification of this map:
$$ 
   \begin{array}{ll}
   \overline{p}\,=\,p\,+\,x\,-\,\frac{1}{2} & mod\ C \\
   \overline{x}\,=\,x\,+\,\overline{p} & mod\ 1 
   \end{array} 
   \eqno (1.2)
$$
where $C\gg 1$ is a circumference of the phase space torus allows for a slow
(diffusive) relaxation with $t_r\sim C^2/4D_p$ where $D_p=1/12$ is 
the diffusion
rate in $p$. A convenient characteristic of BF size is the rms phase space 
volume (area)
$\Gamma (t)=\sigma_p\cdot\sigma_x$ for a group of $N$ trajectories.
In ergodic motion at equilibrium $\Gamma =\Gamma_0=C/12$.
Below I will use the dimensionless measure $\tilde{\Gamma}=\Gamma /\Gamma_0
\to \Gamma$, and omit tilde.

The entropy $S$ can be defined by the relation:
$$
   S(t)\,=\,\ln{\Gamma (t)} \eqno (1.3)
$$
with $S=0$ at equilibrium. This definition is not identical to the standard
one (via the distribution function, or phase space density) but it is fairly
close to the latter if $\Gamma\ll 1$, i.e. for a BF, just what one needs in
the problem under consideration. A great advantage of definition (1.3) is
in that the computation of $S$ does not require very many trajectories
as does the distribution function. In fact, even a single trajectory is
sufficient!

A finite number of trajectories used for calculating the
phase space volume $\Gamma$ is a sort of the coarse--grained distribution,
as required in relation (1.3), but with a free bin size which can be 
arbitrarily small.
The detailed study of BF in this class of ES models will be published 
elsewhere \cite{10}. Here I briefly present just an example shown in Fig.1.

\begin{figure}[]
\centerline{\epsfxsize=15cm \epsfbox{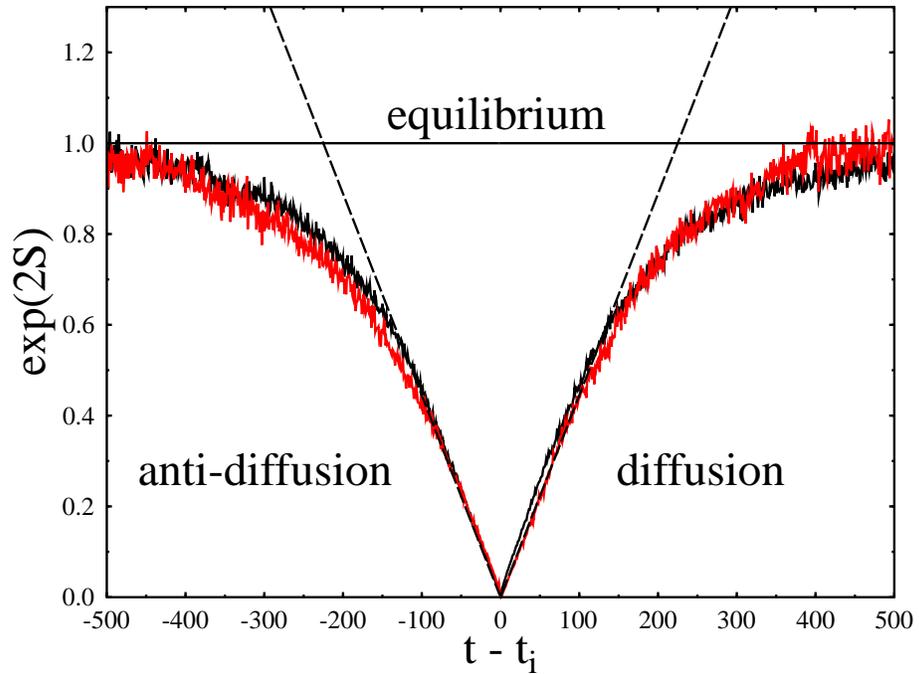}}
\caption{Boltzmann's diffusive fluctuations in model (1.2) with
parameter $C=15$: the square of phase space area
occupied by $N$ independent trajectories ('particles') vs. the time 
(number of map's iterations 
$t-t_i$) counted off the instant $t_i$ of fluctuation maximum,
or of minimal $\Gamma_{fl}$, for each of $N_{fl}$ 
superimposed BF
separated by average period $P=\langle (t_i-t_{i-1})\rangle$.
Straight lines show the expected dependence for anti--diffusion and diffusion
(see text).
Two slightly different curves correspond to $N=1$ (grey) and $N=4$ (black) with
$\Gamma_{fl}=0.0001\ {\rm and}\ 0.1$; $N_{fl}=3352\ {\rm and}\ 2851$; 
$P=29863\ {\rm and}\ 35110$, respectively.
}
\end{figure}

The data were obtained from running 4 and just 1 (!) trajectories for a 
sufficiently
long time in order to collect fairly many BF
which were superimposed in Fig.1 to clean up the regular BF from a 'podlike 
trash' of stationary fluctuations.
The size of BF chosen was approximately fixed by the condition that current
$\Gamma (t)\le\Gamma_{fl}$. In spite of inequality the mean values
$\langle\Gamma (t_i)\rangle =0.000033\ {\rm and}\ 0.069$ are close 
(in order of magnitude) to the 
fixed $\Gamma_{fl}$ values in Fig.1. Notice that for a slow diffusive
kinetics the quantity $\exp{(2S)}\propto\sigma_p^2\propto\langle p^2\rangle$
while $\sigma_x$ remains constant.

The probability of BF can be characterized by the average period
between them for which a very simple estimate
$$
   P\,\approx\,3\,\Gamma_{fl}^{-N}\,\approx\,3\,\exp{(-NS_{fl})}
   \eqno (1.4)
$$
is in a good agreement (upon including the empirical factor 3) 
with data in Fig.1.

In the example presented here the position of all BF in the phase space 
is fixed to
$x_{fl}=1/2,\ p_{fl}=C/2$. If one lifts this restriction the probability
of BF increases by a factor of $1/\Gamma_{fl}$, or by decrease of $N$ by one
($N\to N-1$), due to an arbitrary position of BF in phase space. 
In the former case, a chain of BF are but the well known
Poincar\'e recurrences. What is less known that the latter are a particular
and specific case of BF, and as such the recurrence of a trajectory in chaotic
system is determined by the kinetics of the system.
Recurrence of several $(N>1)$ trajectories can be also interpreted 
as the recurrence
of a single trajectory in $N$ uncoupled freedoms.

As is seen in Fig.1 the irregular deviations from a regular BF are rapidly
decreasing with the entropy $S\to S_{fl}$. One may get the impression that
near BF maximum the motion becomes regular, hence the term 'optimal
fluctuational path' \cite{1}. In fact, the motion remains diffusive down
to the dynamical scale which in model (1.2) is $|\Delta p|\sim 1$ independent
of parameter $C$. 

Big fluctuations are not only perfectly regular by themselves but also
surprisingly stable against any perturbations, both regular and chaotic.
Moreover, the perturbations
do not need to be small. At first glance, it looks very strange
in a chaotic, highly unstable, dynamics. The resolution of
this apparent paradox is in that the dynamical instability of motion
does affect the BF instant of time $t_i$ only. 
As to the BF shape, it is determined by the kinetics whatever its mechanism,
from purely dynamical one, like in model (1.2), to a completely noisy
(stochastic, cf. Fig.1 above and Fig.4 in \cite{1}).
As a matter of fact, the fundamental Kolmogorov theorem \cite{4} is related
just to the latter case but remains valid in a much more general situation.
Surprising stability of BF is similar to the full (less known) property of
robustness of the Anosov (strongly chaotic) systems \cite{11} whose
trajectories get only slightly deformed under a small perturbation 
(for discussion see \cite{12}).
From a different perspective this stability can be interpreted as a fundamental
property of the 'macroscopic' description of BF. 
In such a simple few--freedom system like (1.2) the 'macroscopic' refers
to averaged quantities as $\sigma\,,\ \Gamma\,,\ S$, and the like.
Still, a somewhat confusing result is in that the 'macroscopic' stability
comprises not only the relaxation of BF but also its rise as the both parts
of BF appear always together. It may lead to another misunderstanding that
the probability of fluctuation and relaxation are equal which is certainly
wrong. The point is that the ratio of both (unequal!) probabilities is
determined by a crossover parameter
$$
   R_{cro}(S_{fl})\,=\,\frac{P}{t_r}\,\approx\,\frac{3\,
   \exp{(-NS_{fl})}}{C^2}\,\gg\,1    \eqno (1.5)
$$
Here the latter expression refers to model (1.2), and the inequality
determines the region of BF where
the time of awaiting BF is much longer than 
that of its immediate relaxation
from a nonequilibrium 'macroscopic' state (for further discussion see
Section 6 below).

\section{A new class of dynamical models:\newline What they are for?}
A fairly simple picture of BF in the systems with
equilibrium steady state (ES) is well understood by now, though not yet 
well known. To Boltzmann such a picture was the basis of his fluctuation
hypothesis for our Universe. Again, as is well understood by now such a
hypothesis is completely incompatible with the present structure of the
Universe as it would immediately imply the notorious 'heat death' (see, e.g.,
\cite{13}). For this reason, one may even term such systems the {\it heat
death models}. Nevertheless, they can be and actually are widely used in the description
and study of local statistical processes in {\it thermodynamically} closed
systems. The latter term means the absence of any heat exchange with the
environment. Notice, however, that under conditions of the exponential 
instability of motion 
the only {\it dynamically} closed system is the whole Universe. Particularly, 
this excludes
the hypothetical 'velocity reversal' still popular in debates
over 'irreversibility' since Loschmidt 
(for discussion see, e.g., \cite{12,14} and Section 6 below). 

In any event, dynamical models with ES do not tell us the whole story of
either the Universe or even a typical macroscopic process therein.
The principal solution of this problem, unknown to Boltzmann, is quite
clear by now, namely, the 'equilibriumfree' models are wanted.
Various classes of such models are intensively studied today.
Moreover, the celebrated cosmic microwave background tells us that 
our Universe {\it was born} already in the state of a heat death which, 
however, fortunately to us all became {\it unstable} 
due to the well--known Jeans gravitational instability
\cite{15}. This resulted in developing of a rich variety of collective
processes, or {\it synergetics}, the term recently introduced or, 
better to say, put in use by Haken \cite{16}.
The most important peculiarity of such a collective instability is in that
the total overall relaxation (to somewhere ?) with  ever increasing total 
entropy 
is accompanied by an also increasing phase space {\it inhomogeneity} 
of the system, particularly in temperature.
In other words, the whole system as well as its local parts become 
more and more
{\it nonequilibrium} to the extent of the birth of a {\it secondary} dynamics
which may be, and is sometime, as perfect as, for example, the celestial 
mechanics 
(for general discussion see, e.g., \cite{17,18,12}).

I stress that all these inhomogeneous nonequilibrium structures are not
BF like in ES systems but are a result of regular collective instability,
so that they are immediately formed under a certain condition.
Besides, they are typically {\it dissipative structures} in Prigogine's 
term \cite{19} due to exchange of energy and entropy with the
{\it infinite} environment. The latter is the most important feature of such
processes, and at the same time the main difficulty in studying the
dynamics of those
models both theoretically and in numerical experiments which are so much
simpler for the ES systems. Usually, the investigations in this field 
are based upon statistical laws omitting the underlying 
dynamics from the beginning.

Recently, however, a new class of dynamical models 
has been developed by Evans, Hoover, Morriss, Nos\'e, and others \cite{20,21}.
Some researchers still hope that such brand--new 
models
will help to resolve the 'paradox of irreversibility'. A more serious reason
for studying these models is in that they allow for a fairly simple
inclusion in a few--freedom model the infinitely dimensional 'thermostat',
or 'heat bath'. This greatly facilitates both numerical experiments as well as
the theoretical analysis. Particularly, by using such a model the 
derivation
of Ohm's law was presented in \cite{22}, thus solving 'one of the outstanding
problems of modern physics' \cite{23} (for this peculiar dynamical model only!).
The authors \cite{22} claim that "At present, no general statistical mechanical
theory can predict which microscopic dynamics will yield such transport 
laws..." In my opinion, it would be more correct to inquire which of many 
relevant models could be treated theoretically, and especially in a 
rigorous way as was actually done in \cite{22}.

The zest of new models is the so--called Gauss thermostat, or heat bath (GHB).
In the simplest case the motion equations of a particle in such a bath are
\cite{20,21,22}:
$$
   \frac{d{\bf p}}{dt}\,=\,{\bf F}\,-\,\zeta{\bf p}\,, \qquad
   \zeta\,=\,\frac{{\bf F}\cdot{\bf p}}{p^2} \eqno (2.1)
$$
where ${\bf F}$ is a given external force, and $\zeta$ stands for the 
'friction coefficient'. The first peculiarity of such a 'friction' is in its
explicit time reversibility contrary to the 'standard friction'.
The price for reversibility is the strict connection between the two forces,
the friction and the external force ${\bf F}$. Moreover, 
and this is most important,
the connection is such
that $|{\bf p}|^2=p_0^2=const$ is the exact motion invariant:
$$
   \frac{d}{dt}\,\frac{|{\bf p}|^2}{2}\,=\,{\bf p}\cdot\frac{d{\bf p}}{dt}\,=
   \,{\bf p}\cdot{\bf F}\,-\,{\bf F}\cdot{\bf p} \eqno (2.2)
$$
The first of two identical terms represents the mechanical work of the
external regular force ${\bf F}$, the spring of the external energy, while the 
second one describes the sink of energy into GHB. Thus, asymptotically as
$t\to\infty$ the model describes
a steady state only. This is the main restriction of such models.
The particle itself does only immediately transfer the energy without any
change of its own one due to the above constraint $|{\bf p}|^2=const$.
In one freedom the latter would lead to a trivial solution $p=const$.
So, at least two freedoms are required to allow for a variation 
of vector ${\bf p}$ in spite of constraint. For many {\it interacting}
particles
the constraint $\sum |{\bf p}_i|^2=const$ would be still less, hence 
reference to the Gauss 'Principle of Least Constraint' \cite{24} 
for deriving the reversible friction in Eq.(2.1).
In the present paper the simplest case of $N$ {\it noniteracting}
two--freedom particles is considered only as in \cite{22}.

The next important point is a special form of the energy in GHB which is
the heat.
In true heat bath it would be a chaotic motion of infinitely many particles
therein. This is not the case in GHB, and one needs an additional force
in Eq.(2.1) which would make the particle motion chaotic maintaining,
at the same time, the constraint. Whether such an external to GHB chaos
is equivalent to the chaos inside the true heat bath, at least statistically,
remains an open question but it seems plausible from the physical viewpoint
\cite{22} (see also Ref.\cite{25}). If so, the model describes 
the direct conversion of mechanical work
into heat $Q$, and hence the permanent {\it entropy production}. 
The calculation
of the later is not a trivial question (for discussion see \cite{20,21,22}).
In my opinion, the simplest way is to use the thermodynamic relation:
$$
   \frac{dS}{dt}\,=\,\frac{1}{T}\cdot\frac{dQ}{dt}\,, \qquad
   \frac{dQ}{dt}\,=\,{\bf p}\cdot{\bf F} \eqno (2.3)
$$
where $T=p_0^2$ is an effective 
temperature \cite{22}. 
Since the input energy is of zero entropy (formal temperature 
$T_{in}=\infty$) the relation (2.3) determines the entropy production 
in the {\it whole system} (particles + GHB). Notice that in Eq.(2.3), as well as
throughout this paper, the entropy $S$ is understood as determined
in the standard way via a {\it coarse--grained} phase--space density
(distribution function).

Meanwhile, usual interpretation of GHB models is quite different
\cite{20,21,22}. Namely, the entropy production (2.3) is expressed
via the Lyapunov exponents $\lambda_i$ of particle's motion:
$$
   \frac{dS}{dt}\,\equiv\,\frac{dS_{GHB}}{dt}\,\equiv\,-\frac{dS_p}{dt}\,=\,
   -\sum_i\,\lambda_i \eqno (2.4)
$$
where $S_{GHB}$ and $S_p$ is the entropy of GHB and  of the ensemble of 
particles,
respectively. An unpleasant feature of this relation is in that the latter
equality holds true for the Gibbs entropy only which is conserved in 
Hamiltonian
system modeled by the GHB. As a result the entropy of the total system
(particles + GHB) remains constant (the second equality in Eq.(2.4)) which
literally means no entropy production at all!
Even though such an interpretation can be formally justified it seems to me
physically misleading.
In my opinion, the application of Lyapunov exponents would be better 
restricted to characterization of the phase--space fractal microstructure of 
particle's
motion (which is really interesting) retaining the universal coarse--grained 
definition of the entropy 
(cf. ES models in Section 1). 

As was already mentioned above the GHB models describe the nonequilibrium
{\it steady states} only. Moreover, any collective processes of interacting
particles are also excluded, just those responsible for the very existence
of regular nonequilibrium processes, particularly, of the field ${\bf F}$ in
model (2.1). In a more complicated Nos\'e - Hoover version of GHB models
these severe restrictions can be partly, but not completely, lifted.
Whether it would be sufficient for the inclusion of collective processes
remains, to my knowledge, an open question.

In any event, even the simplest GHB model like (2.1) represents a qualitatively
different type of statistical behavior as compared to that in the ES models.
The origin of this principal difference is twofold:
(i) the external 'inexhaustible' spring of energy, if only introduced
    'by hand', and
(ii) a heat sink of infinite capacity which excludes any equilibrium.

In conclusion of this Section I formulate precisely the model to be
considered below, in the main part of the paper.
Choosing the model for numerical experiments I follow my favored
'golden rule': construct the model as simple as possible but not simpler.
In the problem under consideration the models studied already are mainly 
based on
a well--known and well--studied 'Lorentz gas' that is
a particle (or many particles) moving through a set of fixed scatterers.
A new element is a constant field accelerating particles. Actually, the
Lorentz model becomes in this way the famous Galton Board \cite {26},
the very first
model of chaotic motion, which had been invented by Galton for another purpose,
and which
has not been studied in detail until recently \cite{20,21,22}.
My model is still simpler, and is specified by the two maps:
(i) the 2D Arnold cat map (1.1) to chaotize particles, and
(ii) 1D--map version of Eq.(2.1):
$$
  \overline{p_1}\,=\,p_1\,+\,F\,-\,4Fp_1^2 \eqno (2.5)
$$
where $p_1=p-p_0$, and parameter in Eq.(2.1) $p_0=1/2$. For $|F|<1/4$
the momentum $p$ remains within the unit interval ($0\leq p <1$) as in
map (1.1). The principal relation (2.3) for the entropy reduces now also to
the additional 1D map:
$$
   \overline{S}\,=\,S\,+(p_1\,+\,F)^2\,-\,p_1^2\,=\,S\,+\,2p_1F\,+\,F^2
   \eqno (2.6)   
$$
where the entropy unit is changed by a factor of 2 for simplicity.
Since $S$ is the entropy produced in GHB the latter map implicitly includes
also 
the motion in the second freedom for each of noninteracting
particles due to the Gauss constraint which guarantes the immediate transfer
of energy to GHB.

In numerical experiments considered below an arbitrary number $N$ of
{\it noninteracting} particles (trajectories) with random initial conditions
were used. In this case, the Gauss constraint remained unchanged, and all
trajectories were run simultaneously.

\section{Nonmonotonic entropy production:\newline
Local fluctuations}
Statistical properties of the entropy growth in the model chosen are
determined by the two first moments of the $p_1$ distribution function.
In the limit $t\to\infty$ and/or $N\to\infty$ they are
(per iteration and per trajectory):
$$
   \langle p_1\rangle\,=\,0\,, \qquad  \langle p_1^2\rangle\,=\,
   \frac{1}{12} \eqno (3.1)
$$
where averaging is done over both the motion time $t$ (now the number of map's
iterations), and $N$ noninteracting particles (particle's trajectories).
In combination with Eq.(2.6) the first moment in Eq.(3.1) implies
the linear growth of the average entropy (per trajectory):
$$
   \langle S(t)\rangle\,=\,t\,F^2 \eqno (3.2)
$$

In this Section the statistics of local fluctuations is considered.
A similar problem was studied in \cite{27} for a more realistic model
with many interacting particles. In the present model the local fluctuation
is defined as follows. The total motion time $t_f$ is subdivided into many 
segments of equal duration $t_1$. On each segment $i=1,...,t_f/t_1$ the total 
change
of entropy $S_i$ for all $N$ trajectories is calculated using Eq.(2.6) and 
represented as a dimensionless random variable
$$
   S_{\sigma}\,=\,\frac{S_i\,-\,\langle S_i\rangle}{\sigma}\,=\,
   \frac{S_i\,-\,\tau}{\sigma} \eqno (3.3)
$$
where $\langle S_i\rangle =Nt_1F^2=\tau$
(see Eq.(3.2)), and the rms fluctuation $\sigma$ 
is given by a simple
relation (see Eqs.(2.6) and (3.1)):
$$
   \sigma^2\,=\,\frac{\tau}{3}  \eqno (3.4)
$$
This relation neglects all the correlations which implies the standard
Gaussian distribution:
$$
   G(S_{\sigma}) \,=\,\frac{\exp{\left(-\frac{S_{\sigma}^2}{2}\right)}}
   {\sqrt{2\pi}} \eqno (3.5)
$$

An example of actual distribution function is shown in Fig.2 for
a single trajectory with segment length $t_1=10,\,25,\,100$ iterations,
and the number of segments up to $10^7$.
A cap of the distribution is close to the standard Gauss (3.5) 
(see also Fig.3) but
both tails clearly show a considerable enhancement of fluctuations
depending on both $t_1$ and $N$ (in other examples, see below).

\begin{figure}[]
\centerline{\epsfxsize=15cm \epsfbox{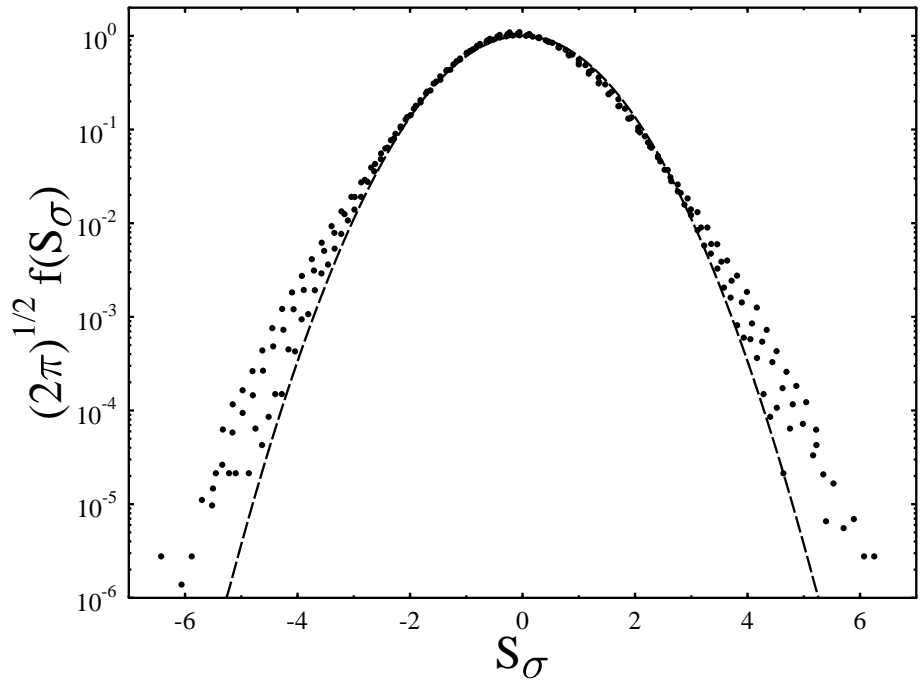}}
\caption{Distribution function $f(S_{\sigma })$ of local fluctuations
in nonequilibrium steady state with $F=0.01$. Dashed line is standard Gauss 
(3.5);
points represent the results of numerical experiments with $N=1$,
and $t_1=10,\,25,\,100$
}
\end{figure}

The shape of the tails is also Gaussian but the width is the larger the
smaller $t_1$ and $N$. This is especially clear in a different representation
of the data in Fig.3 where the ratio of empirical distribution to standard
Gauss is plotted as a function of Gaussian variable $S_G=S_{\sigma}^2/2$.
\begin{figure}[]
\centerline{\epsfxsize=15cm \epsfbox{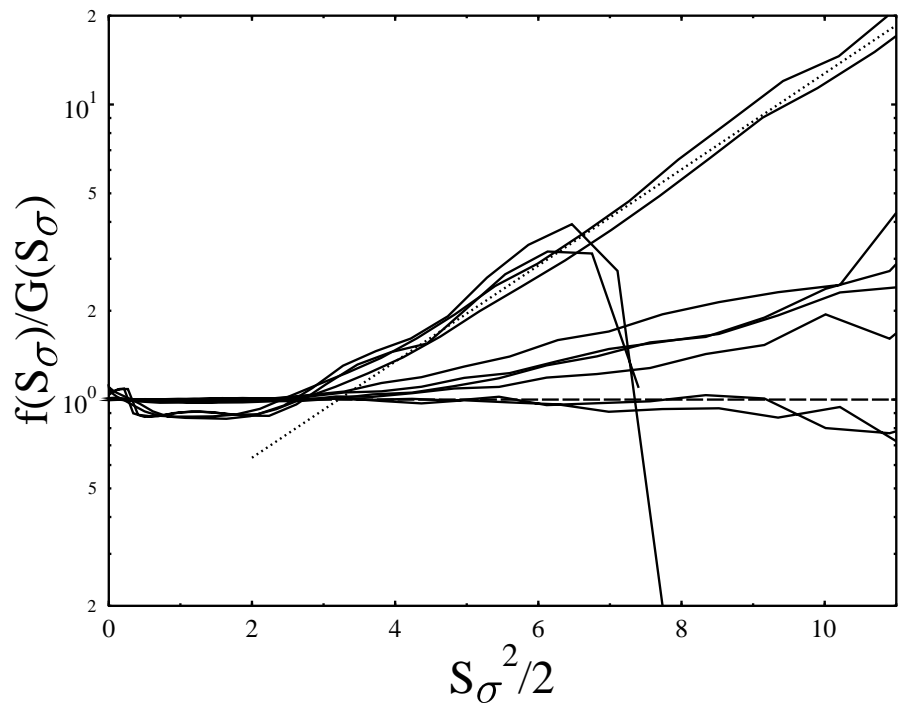}}
\caption{The ratio of distribution $f(S_{\sigma})$ 
to the standard Gauss (3.5) (broken lines). The values of parameter 
$N/t_1$ from 
top to bottom are: 1/5 ($S_{\sigma}^2/2<7.5$, see text); 1/10; 1/100; 10/10, 
and
100/1. The oblique dotted straight line demonstrates the Gaussian 
shape of the tails.
}
\end{figure}
Each run with particular values of $N$ and $t_1$ is represented by the two
slightly different lines for both signs of $S_{\sigma}$. Besides fluctuations
the difference apparently includes some asymmetry of the distribution with
respect to $S_{\sigma}=0$. The origin of this asymmetry 
is not completely clear as yet.
A sharp crossover between the two Gaussian distributions at
$S_G\approx 3$ is nearly independent 
of the parameters
$N$ and $t_1$ as is the top distribution below crossover.
To the contrary, the tail distribution essentially
depends on both parameters in a rather complicated way.
The origin of the difference between the two Gaussian distributions
apparently lies in dynamical correlations. In spite of a fast decay
(see Section 1) the correlation in Arnold map (1.1) does affect somehow
the big entropy fluctuations except the limiting case $N\gg t_1$
(two lower lines in Fig.3) when correlations vanish because of random and
statistically independent initial conditions of many trajectories.

For any fixed parameters $N$ and $t_1$ the fluctuations are 
bounded ($F\ll 1$):
$$
   |S_{\sigma}|\,<\,\sqrt{3Nt_1} \eqno (3.6)
$$
which follows from Eqs.(2.6), (3.3), and (3.4). This is clearly seen in Fig.3
for minimal $Nt_1=5$. If instead only force $F$ is fixed, the relative entropy
fluctuations
$$
   \frac{S_i}{\langle S_i\rangle}\,\approx\,\pm\,\frac{1}{F} \eqno (3.7)
$$
are also restricted but can be arbitrarily large for small $F$ and, moreover,
of both signs. This implies a {\it nonmonotonic} growth of entropy 
at the expense of the segments with $S_i<0$.

The probability (in the number of trajectory segments) of extremely large
fluctuations, Eqs.(3.6) and (3.7), is exponentially small (see Eq.(3.5)
and below).
However, the probability of the fluctuations with {\it negative} entropy
change ($S_i<0$) (without time reversal!)
is generally not small at all, reaching 50\% as $\tau\to 0$ (for arbitrary
$N$ and $t_1$). In principle, it is known, at least for the systems with
equilibrium steady state (Section 1). Nevertheless, the first, to my 
knowledge, direct observation of this phenomenon in a nonequilibrium 
steady state \cite{27} has so much
staggered the authors that they even entitled the paper 'Probability of
Second Law violations in shearing steady state'. In fact, this is simply 
a sort of peculiar fluctuations,
big ones not so much with respect of their size but primarily of their 
probability
(cf. discussion in Section 1). However, the important point is that 
all those
negative entropy fluctuations (transforming the heat into work) are 
randomly scattered among the others of positive entropy, 
and for making any use 
of the former
a Maxwell's demon is required who is known by now to be well in 
a 'peaceful coexistence' 
with the Second Law.

Another interesting limit is $t_1\to t_f\to\infty$ (a single segment) \cite{27}
while $\tau\to 0$ which is possible if $F\to 0$ too.
In this case the probability of {\it zero} entropy change in the whole motion
is also approaches 50\%. However, the probability of {\it any negative}
entropy fluctuation vanishes (see Eq.(3.3)). 
An interesting question is whether there exists
some intermediate region of parameters where the latter
probability would remain finite.
In other words, are the Poincar\'e recurrences to negative
entropy change $S_i<0$ possible in a
nonequilibrium steady state as they are in the equilibrium (Section 1)?
The answer to this question is given by the statistics of the global 
fluctuations.

\section{Nonmonotonic entropy production:\newline
Global fluctuations}
The definition of the global fluctuations is similar to, 
yet essentially different
from, that of the local fluctuations in the previous Section. Namely (cf. Eqs.(3.3)
and (3.4)), the principal dimensionless random variable $S_{\sigma}(t)$ 
now explicitly depends on time:
$$
   S_{\sigma}(t)\,=\,\frac{S(t)\,-\,\langle S(t)\rangle}{\sigma}\,=\,
   \frac{S(t)\,-\,\tau}{\sigma} \eqno (4.1)
$$
where $S(t)$ is calculated from Eq.(2.6), $S(0)=0$, $\langle S(t)\rangle =
NtF^2\equiv\tau$
(see Eq.(3.2)), and the rms fluctuation $\sigma$ 
is given by the same relation (3.4) with a new time variable $\tau$:
$$
   \sigma^2\,=\,\frac{\tau}{3}   \eqno (4.2)
$$
In other words, the global fluctuations are described as a diffusion
with the constant rate:
$$
   D\,=\,\frac{\sigma^2}{\tau}\,=\,\frac{1}{3} \eqno (4.3)
$$

Also, one can view the global fluctuations as a continuous time--dependent
deviation of the entropy from its average growth unlike the local fluctuations
in the ensemble of fixed trajectory segments (Section 3).
Now, the primary goal is to find out if the entropy can reach the negative
values $S(t)<0$ for $t\to\infty$. As was discussed in the previous Section
this is possible at some finite segments of trajectory with the probability
rapidly decreasing (but always finite) as the segment length grows.

In Fig.4 the three examples of global fluctuations are shown in a slightly
different representation (cf. Eq.(4.1))
$$
   S_g(\tau )\,=\,\frac{S(\tau )}{\tau}\,-\,1 \eqno (4.4)
$$
in order to always keep before one's eyes the most important border 
$S(\tau )=0$ ($S_g(\tau )=-1$, a horizontal line in Fig.4).
\begin{figure}[]
\centerline{\epsfxsize=15cm \epsfbox{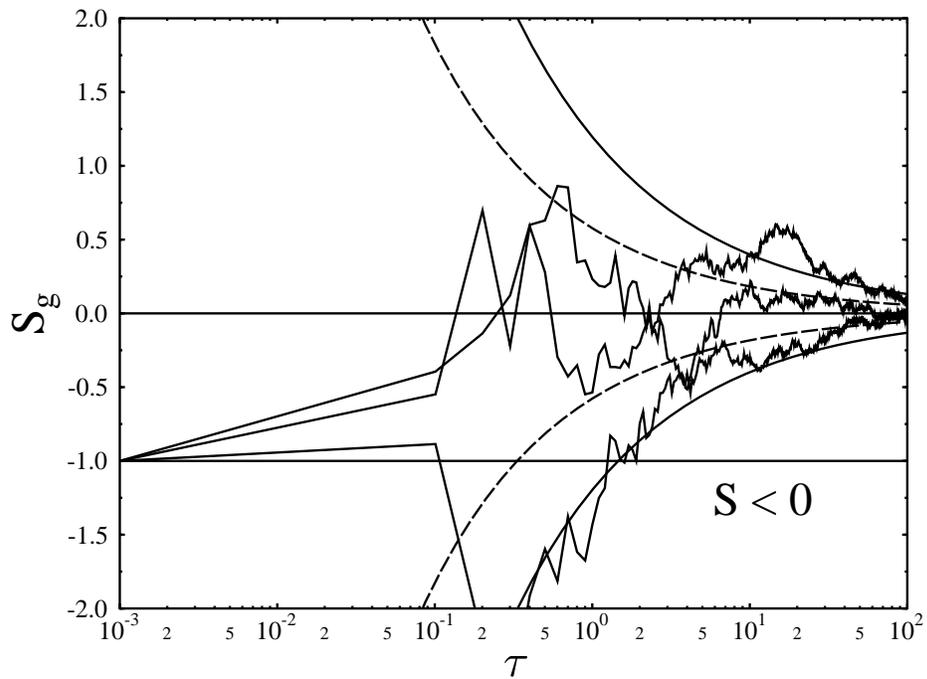}}
\caption{Time dependence of the reduced global fluctuations $S_g(\tau )$,
Eq.(4.4): three sets by $N=10$ trajectories with different initial
conditions but the same initial entropy $S(0)=0$ and $F=0.01$.
Horizontal solid line $S_g=0$ represents the average entropy growth.
The lower solid line $S=0$ is the border between positive and
negative entropy. 
A pair of dashed curves corresponds to the standard rms fluctuation $\sigma$,
Eq.(4.2), and
two solid curves represent the maximal diffusion fluctuations $\sigma_b$, 
Eq.(4.5).
}
\end{figure}
Eventually, all trajectories converge to the average entropy growth 
(a horizontal line $S_g=0$ in Fig.4).
During the initial stage of diffusion the probability of negative entropy
is roughly 50\%, similar to the local fluctuations (Section 3).
However, at $\tau\gsim 1$ the situation cardinally changes, so that all the
trajectories move away from the border $S=0$. Moreover, the relative distance
to the border with respect to the fluctuation size indefinitely increases.

The fluctuation size is characterized by the two parameters.
The first one is the well known rms dispersion $\sigma$, Eq.(4.2) (two dashed
curves in Fig.4), which estimates the fluctuation distribution width.
In the problem under consideration the most important is the second 
characteristic, $\sigma_b$ (two solid curves in Fig.4), which sets the
{\it maximal} size (the upper bound) of the diffusion fluctuations, 
and thus insures
against the recurrence into the region $S<0$ in a sufficiently long time.
The ratio of two sizes
$$
   R_{\sigma}(\tau )\,=\,\frac{\sigma_b}{\sigma}\,=\,\sqrt{2\ln{\ln{(A\tau )}}}
   \eqno (4.5)
$$
is given by the famous Khinchin law of iterated logarithm \cite{28}.

I emphasize again that the principal peculiarity and importance of 
{\it border} $\sigma_b$ is in that it characterizes a {\it sharp drop}
of the fluctuation probability down to {\it zero} (in the limit 
$\tau\to\infty$). In other words, almost any trajectory approaches
{\it infinitely many times}
arbitrarily close to this border from below but the number of crossings
the border remains {\it finite}.
In Fig.4 this corresponds to the eternal confinement
of trajectories in the gap between the two solid curves.

Such a surprising behavior of random trajectories is well known to
mathematicians but, apparently, not to physicists.
In Fig.5 a few examples of the fluctuation distributions are shown for
illustration of that {\it unpenetrable border}.

\begin{figure}[]
\centerline{\epsfxsize=15cm \epsfbox{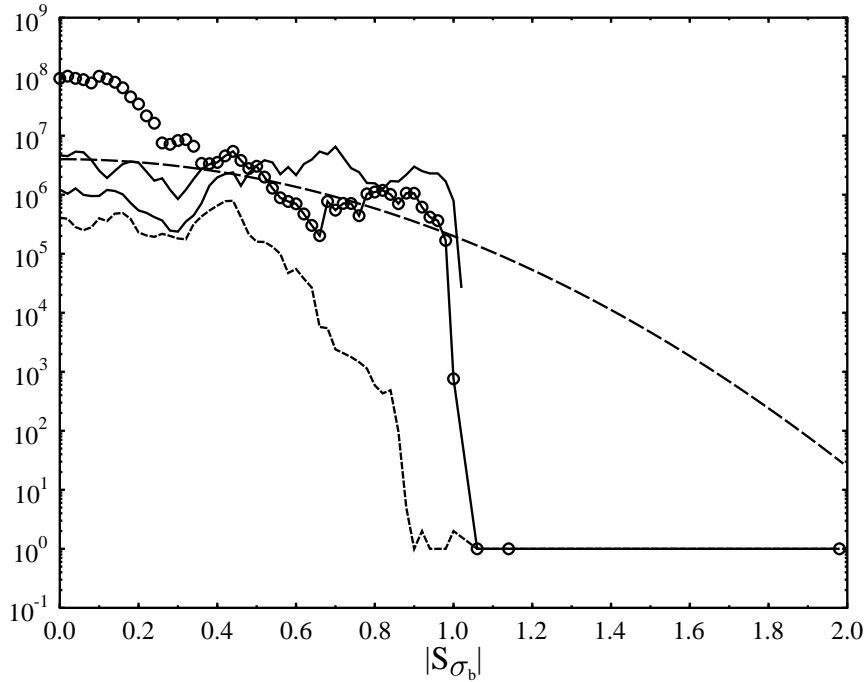}}
\caption{Histogram of the global fluctuations in the number of entries
per bin of width 0.02: $F=0.01;\ N=100;\ R_{\sigma}\approx 3$.
From bottom to top in the leftmost part of figure: $\tau =10^5$ (dashed line);
$10^6$ (two solid lines, different initial conditions); $10^7$ (circles);
the total motion time $t=100\,\tau$ iterations.
For comparison, the smooth dashed line shows unbounded Gaussian distribution 
(4.7) for $\tau =10^6$.
}
\end{figure}

In Khinchin's theorem a factor $A$ in Eq.(4.5) is irrelevant and set to $A=1$.
This is because the theorem can be proved in the formal limit $\tau\to\infty$
only as most of theorems in the probability theory (as well as in the ergodic
theory, by the way). However, in numerical experiments on a finite time,
even if arbitrarily large, one needs a correction to the limit expression.
Besides, it would be desirable to look at the border over the whole motion
down to the dynamical time scale which is determined by the correlation decay.
In the model under consideration it is of the order of relaxation time
$t_r\sim 1$ (see Section 1).
The additional parameter $A$ can be fixed by the condition 
$$
   \sigma_b(\tau_1)\,=\,\sigma (\tau_1)\,, \qquad \tau_1\,=\,NF^2 \eqno (4.6)
$$
for minimal $t=1$ on the dynamical time scale of the diffusion.
Then, Eq.(4.5) implies $A\tau_1=5.2$ which is used in Figs.4 and 5.
The condition assumed is, of course, somewhat arbitrary but the dependence
on $A$ remains extremely weak provided $\tau_1\ll 1$.

The histogram in Fig.5 is given in the absolute numbers of trajectory entries
into bins in order to graphically demonstrate a negligible number of 
exceptional
crossings of the border. The exact formulation of Khinchin's theorem admits
a finite number of crossings in infinite time.
Actually, all those 'exceptions' are concentrated within a relatively short
initial time interval $\tau\sim 1$ (for the accepted $A$ value, see Fig.4).

The distribution of entropy fluctuations between the borders is characterized 
by its own big
fluctuations due to a large time interval ($\sim\tau$) required for 
crossing the distribution region (see Eq.(4.3)). The spectacular
precipice of many orders of magnitude is reminiscent of a 
diffusion
'shock wave' cutting away the Gaussian tail. Unbounded
Gauss is also shown in
Fig.5 by the smooth dashed line. 

In variable $S_{\sigma_b}=S_{\sigma}/R_{\sigma}$ the standard Gauss 
is no longer a stationary distribution (cf. Eq.(3.5)):
$$
   \sqrt{2\pi}\,G(S_{\sigma_b})\,=\,R_{\sigma}(\tau )\cdot
   \exp{\left(-\frac{S_{\sigma_b}^2}{2}R_{\sigma}^2
   (\tau )\right)} \eqno (4.7)
$$
Both the probability density at the border $|S_{\sigma_b}|=1$ as well as
the integral probability beyond that are slowly decreasing 
$\sim 1/\ln{(A\tau )}$. The 'shock wave' decays but still continues to 
'hold back' the trajectories.

Thus, unlike unrestricted entropy fluctuations out of the equilibrium steady
state (Section 1) the strictly restricted fluctuations in the nonequilibrium
steady state get well separated, in a short time, from the region of negative 
entropy, separated in a large excess which keeps growing in time.
In other words, the Poincar\'e recurrences to any negative entropy quickly
and completely disappear leaving the system with ever increasing, even if
nonmonotonically, entropy. 

As nonequlibrium steady state involves a heat bath of infinite phase space
volume (or its nice substitute, the Gauss heat bath), the Poincar\'e 
recurrence theorem is not applicable. Yet, the 'anti--recurrence' theorem
is not generally true either. For example, the entropy repeatedly crosses
the line $S=\tau$ of the average growth in spite of infinite heat bath,
yet it does not so for the line $S=0$ of the initial entropy.

\section{Big entropy fluctuations in critical dynamics}
The strict restriction of the global entropy fluctuations in a nonequilibrium 
steady state considered in the previous Section is a result of the 'normal',
Gaussian, diffusion of the entropy with a constant rate (4.3), and with
the surprising unpenetrable border (4.5). In turn, this is related to a
particular underlying dynamics of the model (1.1) with very strong statistical
properties. Notice that the border (4.5) is of a statistical nature as it
is much less than the maximal dynamical fluctuation (3.7).

However, it is well known by now that, generally, the homogeneous diffusion
can be 'abnormal' in the sense that the diffusion rate does depend on time:
$$
   D(t)\,\sim\,t^{c_D}\,, \qquad -1\,\leq c_D\,\leq 1 \eqno (5.1)
$$
where $c_D$ is the so--called critical diffusion exponent. 
The term 'critical' refers
to a particular class of such systems with a very intricate and specific
structure of the phase space (see, e.g., \cite{29} and references therein).
The 'normal' diffusion corresponds to 
$c_D=0$ while a positive $c_D>0$ represents a superfast diffusion with the
upper bound $c_D=+1$, the maximal diffusion rate possible for a homogeneous
diffusion. The latter is, of course, most interesting case for the problem
under consideration here. A superslow diffusion for a negative $c_D<0$ 
is also possible
with the limit $c_D=-1$ which means the absence of any diffusion 
for $c_D<-1$.
An interesting example of a superslow diffusion
with $c_D=-1/2$ was considered in \cite{30}.
Besides a particular application to the plasma confinement
in magnetic field the example is of a special interest
as this slow diffusion is a result of the time--reversible diffusion
of particles in a chaotic magnetic field.
For other examples and various discussions of abnormal diffusion see 
\cite{31}.  

A number of dynamical models exhibiting the superfast diffusion are known
including the limiting case $c_D=1$ \cite{29,32}.
Interestingly, a simple simulation of abnormal diffusion is possible
by a minor modification of the model under consideration. 
It concerns the additional 1D map (2.6) only
which now becomes: 
$$
   \overline{S}\,=\,S\,+(\,2p_1F\,+\,F^2)\cdot t_s  \eqno (5.2)   
$$
Here the new variable $t_s$ is defined also by a simple relation:
$$
   t_s\,=\,s^{-c_s}\,, \qquad s=1\,-\,2|p_1| \eqno (5.3)
$$
where $s$ is the distance from any of the two borders $p_1=\pm 0.5$
homogeneously distributed within the interval ($0<s<1$).
The quantity $t_s>1$ describes the {\it sticking} of a trajectory in
the 'critical structure' concentrated near $s=0$. Actually, in the model
there is no such a structure, yet the effect of that
is simulated by the 'sticking time' $t_s$ which enhances both the fluctuations
and the average entropy (5.2).
In a sense, such a simulation is similar, in spirit, to that of the Gauss
heat bath. All the properties of that sticking are described by a single
parameter $c_s$, the critical sticking exponent ($0\leq c_s\leq 1$). 
Particularly, it is directly
related to the diffusion exponent $c_D$ (see below).

The statistical properties of abnormal diffusion in this model are determined
by the two first moments of $t_s$ distribution which are directly evaluated
from the above relations as follows. For the first moment it is
$$
   \langle t_s\rangle\,=\,\int_0^1t_s(s)\,ds\,=\,\frac{1}{1\,-\,c_s}\,,
   \qquad c_s\,<\,1 \eqno (5.4a)
$$
and
$$
   \langle t_s\rangle\,\approx\,\ln{\frac{1}{s_1}}\,\approx\,\ln{t}\,,
   \qquad c_s\,=\,1 \eqno (5.4b)
$$
In the latter case the integral diverges, and is determined by the minimal
$s\approx s_1\sim 1/t$ reached over time $t$ which is the total motion time
in {\it map's iterations}. This should be distinguished from the 'physical
time' in a true model of the critical structure
$$
  \widetilde{t}\,\approx\,t\cdot\langle t_s\rangle\,\approx\,\left\{
   \begin{array}{ll}
   \frac{t}{1\,-\,c_s}\,, & c_s\,<\,1 \\
   t\,\ln{t}\,, & c_s\,=\,1 
   \end{array} \right . 
   \eqno (5.5)  
$$
In a similar way the second moment is given by three relations:
$$
   \langle t_s^2\rangle\,=\,\frac{1}{1\,-\,2c_s}\,,
   \qquad c_s\,<\,\frac{1}{2} \eqno (5.6a)
$$
for the normal diffusion,
$$
   \langle t_s^2\rangle\,\approx\,\ln{\frac{1}{s_1}}\,\approx\,\ln{t}\,,
   \qquad c_s\,=\,\frac{1}{2} \eqno (5.6b)
$$
in the critical case, and
$$
   \langle t_s^2\rangle\,\approx\,\frac{s_1^{1-2c_s}}{2c_s\,-\,1}\,\approx\,
   \frac{t^{2c_s-1}}{2c_s\,-\,1}\,,
   \qquad \frac{1}{2}\,<\,c_s\,\leq\,1 \eqno (5.6c)
$$
for the superfast diffusion.

The average entropy production is found from Eq.(5.2):
$$
   \langle S_i\rangle\,=\, NF^2t\langle t_s\rangle\,=\,NF^2\widetilde{t}\,
   \equiv\,\tau \eqno (5.7)
$$
with redefined time variable $\tau$ (cf. Eq.(3.3)).
In this Section the simplest case of a single trajectory ($N=1$) will be
considered only.

Evaluating the superfast diffusion requires a slightly different averaging
$\langle (2p_1t_s)^2\rangle$ (see Eq.(5.2)). Yet it is easily verified
that asymptotically, as $\tau\to\infty$, the difference with respect to 
Eq.(5.6c) vanishes, and one arrives at the following estimate for the
critical rms dispersion $\sigma_{cr}$:
$$
   \frac{\sigma_{cr}^2(\tau )}{B^2}\,=\,\widetilde{t}D(\widetilde{t})\,=\,
   F^2\langle t_s^2\rangle t\,=\,\frac{(1\,-\,c_s)^{2c_s}}{2c_s\,-\,1}\cdot
   \frac{\tau^{2c_s}}{F^{4c_s-2}} \eqno (5.8a)
$$
if $1/2<c_s<1$ (5.6c), and
$$
   \frac{\sigma_{cr}(\tau )}{B}\,=\,\frac{\tau}{F\cdot\ln{(\tau /F^2)}}
   \eqno (5.8b)   
$$
in the most interesting limiting case $c_s=1$. Here empirical factor
$B\sim 1$ accounts for all the approximations in the above relations.

The limit $c_s\to 1$ in Eq.(5.8a) crucially differs from the limiting 
relation (5.8b). The origin of this discrepancy is Eq.(5.4a).
A more accurate evaluation for $c_s\approx 1$ reads:
$$
   \langle t_s\rangle\,=\,\int_{s_1}^1t_s(s)\,ds\,=\,
   \frac{1\,-\,s_1^{1-c_s}}{1\,-\,c_s}\,=\,
   \frac{1\,-\,\exp{[(1-c_s)\ln{s_1}]}}{1\,-\,c_s} \eqno (5.9)
$$
where $s_1\sim 1/t$ is the minimal $s$ over $t$ map's
iterations (cf. Eq.(5.4b)). Hence, the relation (5.4a) is valid
under condition $\epsilon\ln{t}>1$ only ($\epsilon =1-c_s$) 
while in the opposite
limit $\langle t_s\rangle\approx\ln{t}$ as for $c_s=1$, Eq.(5.4b). 
The crossover between the two scalings is at
$$
   t_{cro}\,\sim\,{\rm e}^{1/\epsilon}\,, \qquad \tau_{cro}\,\sim\,
   \frac{{\rm e}^{1/\epsilon}}{\epsilon}\,F^2 \eqno (5.10)
$$
The deviation from Eq.(5.8a) is essential for a sufficiently small $\epsilon$
only. 

The ratio of fluctuations to the average entropy production is given by
the reduced entropy (see Eq.(4.4))
$$
   S_g=\pm \frac{\sigma_{cr}}{\tau}\,\approx\,\pm\frac{B}{F\cdot
   \ln{(\tau /F^2)}} \eqno (5.11)
$$
where the latter expression is estimate (5.8b) for the rms 
fluctuations. They are slowly decreasing with time,
and at $\tau\gsim\tau_0=
F^2\exp{(1/F)}$ the rms line crosses the border $S_g=-1$ of zero entropy.
Afterwards, the
entropy remains mainly positive. To be more correct,
the probability for a trajectory to enter into the region of negative
entropy is systematically decreasing with time, rather slow though.
This is to be compared with the $F$--independent crossover
$\tau_0=1/3$ and a rapid drop of the probability to return back to $S<0$
in case of the normal diffusion (Section 4).

However, there exists another mechanism of big fluctuations, specific for the
critical dynamics. Namely, a separated individual fluctuation can be produced
as a result of a single extremely big sticking time $t_s$ over the
total motion up to the moment the fluctuation springs up in a single map's
iteration.
I remind that in the present model each sticking corresponds to just one map's
iteration. 
The increments of dynamical variables in such a jump are obtained from
Eq.(5.2):
$$
   \Delta S\,=\,\pm Ft_s\,, \qquad \Delta \tau\,=\,F^2t_s \eqno (5.12)
$$
where $t_s\gg 1\ (2p_1\approx 1)$ is assumed (a big fluctuation).
Then, the reduced fluctuation
$$
   S_g\,\approx\,\frac{S}{\tau}\,=\,\pm\frac{Ft_s}{\tau\,+\,F^2t_s}\,
   \approx\,\pm\frac{1/F}{1\,+\,\frac{\tau}{\Delta\tau}} \eqno (5.13)
$$
The maximal single sticking time over the motion time $t$ is, on the average
$$
   \langle t_s\rangle\,\approx\,t\,\ln{t}\,=\,\frac{\tau}{F^2} \eqno (5.14)
$$
Hence, a single fluctuation (5.13) has the upper bound
$$
   |S_g|\,\lsim\,\frac{A}{F} \eqno (5.15)
$$
Here an empirical factor $A\sim 1$ is introduced
similar to Eq.(5.8b). 

The border (5.15) considerably exceeds the rms diffusion fluctuation (5.11)
and, what is even more important, the
former never crosses the line of zero entropy $S_g=-1$.
Thus, the critical fluctuations repeatedly bring the system
into the region of negative entropy. This is because the upper bound (5.15)
does not depend on time $\tau$ provided $\Delta\tau\gsim\tau$ in Eq.(5.13).
However, in a chain of successive fluctuations the values of $\tau$
in Eqs.(5.13) and (5.14) are not generally equal. While in the
former relation it is always the total motion time as
assumed above, in Eq.(5.14) it should be the preceeding period of
fluctuations:
$\tau_n\to 
P_n<\tau_n$ where $n$ is the serial number of
fluctuations. Hence, the approach to the upper bound (5.15)
is only possible under condition $P_n\gg P_{n-1}$ which implies
$P_n\approx\tau_n$. Thus, the fluctuations become more and more rare
with a period growing exponentially in time. In other words,
the fluctuations are stationary in $\ln{\tau}$ with a quite big mean period
$\langle\ln{P}\rangle\approx 5$ (see Fig.6). 

In Fig.6 an example of a few big critical fluctuations in the limiting case 
$c_s=1$
is presented for five single fairly long trajectories with different
initial conditions, and  the motion time
up to $\tau\approx 5\times 10^9$ and $t=10^{10}$ iterations.
To achieve such a long time the force was 
decreased down to $F=0.1$ (see Eq.(5.14)).
\begin{figure}[]
\centerline{\epsfxsize=15cm \epsfbox{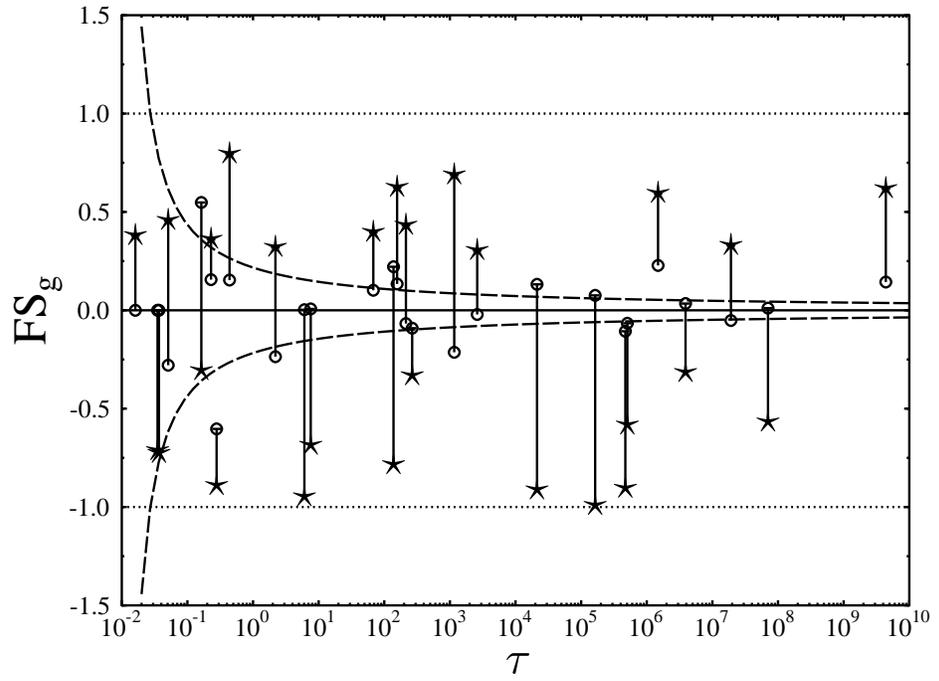}}
\caption{Time dependence of 26 big fluctuations in critical dynamics: 
5 single trajectories up to $10^{10}$ iterations, $c_s=1,\ F=0.1$. 
Only fluctuations with $F|S_g|>0.3$ are shown, each one by a pair of points
connected by the straight line:
the big fluctuation itself (stars), and that at the preceding map's
iteration (circles, see text).
Two dashed curves
show the rms fluctuations of $F|S_g|$, Eq.(5.11) with $B=1$. Horizontal dotted 
lines mark the upper bound, Eq.(5.15) with $A=1$.
}
\end{figure}

Unlike a similar Fig.4 for the normal diffusion, only a few big fluctuations
with $F|S_g|>0.3$ are presented in Fig.6. For a full picture of critical
fluctuations the required output becomes formidably long.
The distribution of all fluctuations, independent of time, is shown below
in Fig.7.

Each fluctuation in Fig.6 is presented by a pair of $FS_g$ values
connected by the straight line:
one at map's iteration just before the fluctuation (circles), and the other
one (stars) at the next iteration when the fluctuation springs up (see above).
Both are plotted at the same, latter, $\tau$ to follow the pairs.
This somewhat shifts the circles to the right.

The most important, if only preliminary, result of numerical experiments
is confirmation of the fluctuation upper bound (5.15)
independent of time. As expected, the circles represent considerably smaller
$F|S_g|$ values roughly following the diffusive scaling (5.11). 

The border (5.15) qualitatively reminds the strict upper bound for the
normal diffusion (Section 4), including a logarithmic ratio with respect
to the rms size (4.5), as compared to the ratio
$$
   R_{cr}(\tau )\,\approx\,\ln{(\tau /F^2)} \eqno (5.16)
$$
in the critical diffusion. An interesting question if the new, critical,
border is also as strict as the old one in the normal diffusion remains, 
to my knowledge, open, at least for the physical model under consideration
where the superdiffusion is caused by a strong long--term correlation 
of successive entropy changes due to the sticking of trajectory.

However, for a much simpler problem of 
statistically independent changes various generalizations of Khinchin's theorem
to the abnormal diffusion were proved by many mathematicians
(see, e.g., \cite{33}).
In the present model it is just the case for the description in map's time
$t$ with statistically independent iterations.
The most general and complete result has been recently obtained by Borovkov 
\cite{baa}.
In the present notations it
can be approximately represented in a very simple form for the ratio (5.16):
$$
   R_{cr}\,=\,{\sigma_b\over\sigma}\,\sim\,(\ln{t})^{c_s}\,\approx\,
   \ln{(\tau /F^2)} \eqno (5.17)   
$$
in the whole interval ($1/2<c_s\leq 1$) of the superdiffusion where the 
physical time $\tau$ is determined by Eq.(5.7). 
Then, for most important reduced fluctuation (5.13)
one arrives at the two relations
$$
   |S_g|\,\lsim\,{\sigma_b\over\tau}\,\sim\,
   \frac{\tau^{c_s-1}}{F^{2c_s-1}}\cdot\left(\ln{{\tau\over F^2}}
   \right)^{c_s} \eqno (5.18a)
$$
for $c_s<1$, and   
$$
   |S_g|\,\lsim\,{\sigma_b\over\tau}\,\sim\, {1\over F} \eqno (5.18b)
$$
in the limiting case $c_s=1$.   
The latter confirms estimate (5.15) above  which,
in turn, is in a good agreement with the empirical data
in Fig.6. 
In any event,
a simple physical estimate (5.15) seems to
provide an efficient
description of
the fluctuation upper bound.

In Fig.7 an example of all (at each map's iteration) fluctuations
is shown
for the data from the same runs as in Fig.6. Besides very large overall 
distribution
\begin{figure}[]
\centerline{\epsfxsize=15cm \epsfbox{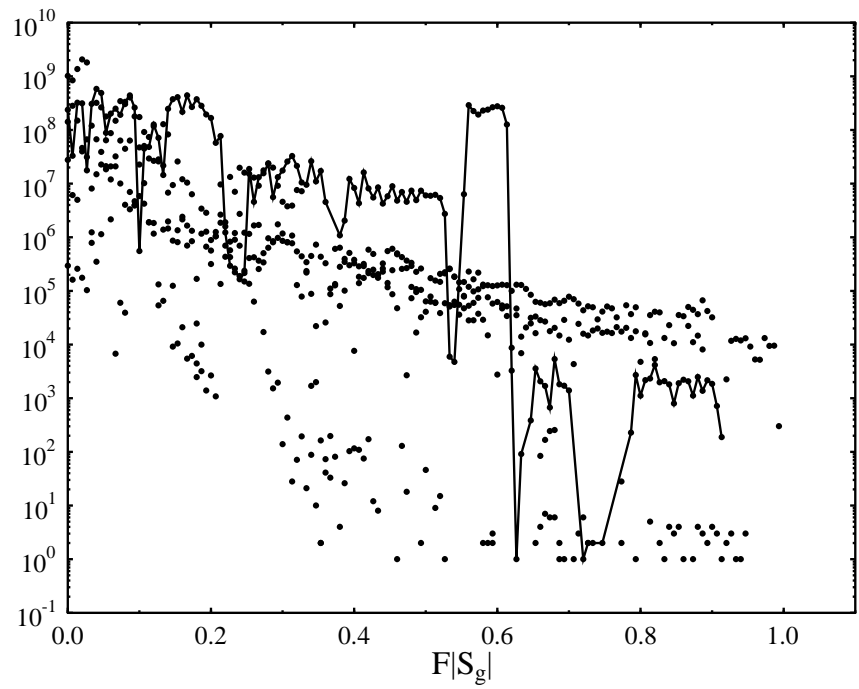}}
\caption{Histogram of critical fluctuations in the number of entries
per bin of width 0.007 for the data in Fig.6. The border $S=0$ corresponds
to $FS_g=-F=-0.1$. The points for the longest trajectory are connected by line.
}
\end{figure}
fluctuations a sharp drop by about four orders of magnitude is clearly
seen near the expected upper bound (5.15). It looks similar to the drop
in Fig.5 for the normal diffusion.

Thus, the critical diffusion results in infinitely many recurrences
far into the region of negative entropy $S<0$ (for $F\ll 1$), 
the sojourn time over there
being comparable to the total motion time. Of course, the former is less
than 50\% on an average, so that asymptotically in time the entropy keeps 
always growing. In this respect, the global critical fluctuations are similar
to the local ones in the normal diffusion (Section 3).

Notice, however, that the upper bound 
$\sigma_b/\tau\sim 1/F$ (5.18b) is permanent in the strict limit $c_s=1$ only.
For any deviation from this limit $\epsilon =1-c_s>0$
this bound lasts a finite time determined by the
crossover (5.10) ($\tau\lsim F^2\exp{(1/\epsilon )}/\epsilon$)
to decreasing $\sigma_b/\tau\to 0$ (5.18a).
Another interesting representation of such an intermediate
behavior is the crossover in the sticking
exponent:
$$
   \epsilon\,\lsim\,{1\over\ln{(\tau /F^2)}}\,\approx\,
   F|S_g| \eqno (5.19)
$$
which is actually shown  in Fig.6 by the upper dashed line.
For the longest $\tau =5\times 10^9$ the latter crossover
$\epsilon_{cro}\approx 0.037$.

\section{Discussion and conclusion}
In the present paper the results of extensive numerical experiments on
big entropy fluctuations in a nonequilibrium steady state of classical
dynamical systems are presented,
and their peculiarities are analysed and discussed. For comparison,
some similar results for the equilibrium steady state are briefly described
in the Introduction (the latter will be published in detail elsewhere
\cite{10}).

All numerical experiments have been carried out on the basis of a very simple
model - the Arnold cat map (1.1) on a unit torus - with only three minor, 
but important,
modifications which allowed for comprising all the problems under consideration
above. The modifications are:

(1) Expansion of the torus in $p$ direction (1.2) which allows for more
    impressive diffusive fluctuations out of the equilibrium steady state 
    (Fig.1 in Section 1).
    
(2) Addition of a 1D map (2.5) with the constant driving force $F$, and
    with ingenious time--reversible friction force which represents the 
    so--called Gauss heat bath, and which allows for modeling a physical
    thermostat of infinitely many freedoms \cite{20,21}.
    This modification is the principal one in the present studies of 
    fluctuations in a nonequilibrium steady state (Sections 3 - 5).
    
(3) Addition of a new parameter $t_s$ (5.3) in map (5.2) which allows for the study of very unusual
    fluctuations of an 'abnormal', critical, dynamical diffusion (Section 5).
    
Big fluctuations in an equilibrium steady state (ES) are briefly considered
in Section 1. The simplest one of this class, which I call the Boltzmann
fluctuation, is shown in Fig.1. It is obviously symmetric with respect
to time reversal, so that at least in this case there is no physical reason 
at all for the conception
of the notorious '{\it time} arrow'. 
Nevertheless, a related conception, say, {\it thermodynamic} arrow, pointing
in the direction of the average increase of entropy, makes sense in spite
of the time symmetry. The point is that the relaxation time of the fluctuation
is determined by model's parameter $C$ only, and does not depend on the
fluctuation itself. On the contrary, the expectation time for a given 
fluctuation, or the mean period between successive fluctuations, rapidly grows 
with the fluctuation size and with the number of trajectories (or freedoms).

Besides the simplest Boltzmann fluctuation, various others are also 
possible, typically with a much less probability.
One of those - the two correlated Boltzmann fluctuations, which I call
the Schulman fluctuation - was recently described in \cite{35} using the same
Arnold cat map. However, such a model has nothing to do with cosmology
as was speculated in \cite{35}. At least, the Universe we live in as well as
most macroscopic phenomena therein require the qualitatively different
models, ones {\it without} equilibrium steady state.
Such structures do appear (with probability 1) as a result of certain regular 
collective processes which lead to very complicated nonequilibrium and 
inhomogeneous states
with ever increasing entropy. This is in contrast
with a constant, on an average, entropy in ES systems. 

A nonequilibrium {\it steady state}, the main subject of the present studies,
is but a little, characteristic though, piece of the chaotic collective 
processes. In the model (2.5) the driving force $F$ represents
a result of some preceding collective processes, the spring of free energy,
while the Gauss friction does so for an infinite environment around, the sink
of the energy, converting the work into heat, on the average.
An interesting peculiarity of such systems is in that the big fluctuations
may do, and do indeed under certain conditions, the opposite, converting back
some heat into the work.

Two types of fluctuations were studied:

(i) the local ones on a set of trajectory segments of length $t_1$ iterations
    and of entropy change $S_i$ (Section 3), and
    
(ii) ones of the global entropy $S(t)$ along a trajectory with respect
     to the initial entropy set to zero: $S(0)=0$ (Sections 4 and 5).
     
The former were found to have a stationary unrestricted distribution close
to the standard Gauss with some enhancement of unknown mechanism
for large fluctuations. The study of the latter effect will be continued.
The distribution is symmetric with respect to the average entropy, growing
in proportion to time in agreement with previous studies on a more complicated
(and more realistic) model \cite{27}. 
Even though the distribution is asymmetric with respect to zero 
entropy change, the probability of negative $S_i<0$ is generally not small
provided $F^2\,Nt_1\lsim 1$. This phenomenon, apparently a new one in the
nonequilibrium steady state, was first observed in \cite{27} but has been interpreted
there as a violation of the Second Law. It seems to be
the reflection
of a common, but wrong in my opinion, understanding the Second Law as a
{\it monotonic} growth of the entropy, thus neglecting all the fluctuations
including the large ones. Meanwhile, the {\it nonmonotonic} rise of entropy
is clearly seen, for instance, in Fig.4, and discussed in detail in
Sections 3 and 4.

The behavior of the global entropy is completely different as the data in the
same Fig.4 demonstrate (Section 4). Even though the entropy evolution remains 
nonmonotonic
it quickly crosses the line of the initial zero entropy and does not return back into the
region of negative entropy $S<0$. This is insured by the famous Khinchin
theorem about the strict upper bound for the diffusion process.
At least for physicists, such a limitation of statistical nature 
for a random motion is surprising and apparently less known. 
That unidirectional evolution is the most important distinction of the
nonequilibrium steady states from the equilibrium ones.

This farther justifies the conception of the {\it thermodynamic arrow} pointing
to a larger, on the everage, entropy. Yet, again it has nothing to do with
the properties of time. Of course, upon formal time reversal the entropy
will systematically decrease, also in the model under consideration because
the Gauss heat bath is time reversible. Within the {\it steady state}
approximarion, or rather restriction, this would be an infinitely large
fluctuation which never came to the end. However, such a fluctuation would 
never occur either, as a result of the natural time evolution of the system,
opposite to the case of equilibrium fluctuations.
The ultimate origin of that crucial difference is in that the former process,
even asymptotically in time, is a tiny little part of the full underlying
dynamics of an infinite system. Particularly, the initial state $S(0)=0$ is
not a result of the preceding fluctuation, as is the case in  ES,
but has been eventually caused, for instance, by instability
of the initial ES at a very remote time in the past. If one would imagine the
time reversal at that instant nothing were changed as the thermodynamic arrow
does not depend on the direction of time provided, of course, the time
reversible fundamental dynamics. That is just this universal overall dynamics
which unifies the time for all the interacting objects like particles and
fields throughout the Universe. Particularly, it is incompatible with the
two opposite time arrows - an old Boltzmann's hypothesis \cite{2} -
which still has some adherents \cite{35}.

Coming back to a particular phenomenon of nonequilibrium steady states
it is worth to mention that the regularities of the fluctuations in those,
both local and global, can be applied, at least
qualitatively, to a small part of a big fluctuation in a statistical
equilibrium (Fig.1) on both sides of the maximum. This interesting question
will be considered in detail elsewhere \cite{10}. 

Finally, some preliminary numerical experiments on the global entropy 
fluctuations 
and the theoretical analysis
were carred out in a special case of the critical dynamics which turned out
to be the most interesting one for the problem in question (Section 5).
The point is that the critical dynamics leads to an 'abnormal' superdiffusion
with the rate $D\propto\tau^{2c_s-1}$ and rms fluctuation size $\sigma_{cr}
\propto\tau^{c_s}$ where $c_s$ is a new parameter of the third model
($1/2<c_s\leq1$). This implies the reduced entropy $|S_g|\propto\tau^{c_s-1}$
decreasing very slowly for $c_s\approx 1$ as compared to the normal diffusion
$|S_g|\propto 1/\sqrt{\tau}$. In the limiting case $c_s=1$ the entropy
$|S_g|\propto 1/\ln{\tau}$ is still decreasing. However, besides diffusive
fluctuations there is a set of infinitely many separated fluctuations whose
size does not decrease with time at all (Fig.6). In other words, these
preliminary numerical experiments suggest that in the limiting case of the
critical dynamics the Poincar\'e recurrences to the initial state $S=0$ and
beyond keep to repeatedly occur without limit. 
These are preliminary results to be
confirmed and farther studied in detail.

In the present studies the fluctuations in classical mechanics 
are presented
only. Generally, the quantum fluctuations would be rather 
different. However, according to the Correspondence Principle, the
dynamics and statistics of a quantum system in quasiclassics are close
to the classical ones on the appropriate time scales of which 
the longest one corresponds just to the diffusive kinetics
providing
transition to the classical limit (for details see \cite{12,34}).
Interestingly, the computer classical dynamics that is the simulation
of a classical dynamical system on digital computer is of a qualitatively 
similar character. This is because any quantity in computer representation
is discrete ('overquantized'). As a result the correspondence between
the classical continuous dynamics and its computer representation in
numerical experiments is restricted to certain finite time scales
like in the quantum mechanics (see two first references \cite{34}).

Discreteness of computer phase space leads to another peculiar phenomenon:
generally, the computer dynamics is irreversible due to the rounding--off
operation unless the special algorithm is used in numerical experiments.
Nevertheless, this does not affect the statistical properties of chaotic
computer dynamics. Particularly, the statistical laws in computer 
representation remain time--reversible in spite of (nondissipative) 
irreversibility of the
underlying dynamics.
This simple example demonstrates that, contrary to a common belief,
the statistical reversibility is a more general property than the dynamical
one.

In the very conclusion I would like to make a brief remark on a very difficult,
complicated and vague problem - the so--called (physical)
causality principle that is
the time--ordering of the cause and effect.
Discussion of this important problem in detail will be published elsewhere
\cite{36}. Here I mention, as an example, a simple Boltzmann's fluctuation
shown in Fig.1. I adhere to the idea of statistical nature of causality.
Indeed, the cause is, by definition, an 'absolutely' independent event
which is only possible in the {\it chaotic dynamics}. Moreover, in any purely 
dynamical description the conception of cause loses its usual physical
meaning. For example, the initial conditions do precisely determine 
the whole infinite
trajectory ($-\infty <t<\infty$) that is both the future as well as the past
of such a 'cause'. In case of a single Boltzmann's fluctuation
an appropriate cause would be the minimal entropy (at $t=t_i$ in Fig.1).
This was exactly the procedure used in numerical experiments for the 
location of a fluctuation of an {\it approximately} given size. 
The principal difference from the exact dynamical initial conditions  
is in that the former cause is an {\it approximate} (e.g., average)
fluctuation's size which is quite sufficient for the complete {\it statistical}
description of the fluctuation, yet does leave behind enough freedom for
independence from other events like preceding fluctuations.
However, this cause does determine not only the future relaxation
of the fluctuation (in agreement with the causality principle) but also
the past rise of the same fluctuation which is a violation of causality,
or acausality (spontaneous rise of a fluctuation), or {\it anti--causality} 
which the latter 
is perhaps the most appropriate
term. Upon the time reversal, the causality/anticausality exchange which
allows for a conception of the {\it causality arrow} but, again, without any
reference to the time of physics. In such a philosophy the directions of both
arrows, thermodynamic and causal one, do coincide independent of the direction
of time. 
An important point of this philosophy is in that the concept 'arrow' is related
to the {\it interpretation} of a physical phenomenon rather than to the
phenomenon itself. Particularly, a question 'how to fix or maintain the arrow'
\cite{35}
is up to the researcher alone.
In a more complicated Schulman's double fluctuation the mechanism of causality
becomes more interesting \cite{35}, and will be discussed, from a different
point of view, in \cite{36}.

\vspace{3mm}

{\bf Acknowledgments.} I am grateful to Wm. Hoover for attracting my
attention to a new class of highly efficient dynamical models with the
Gauss heat bath, and for stimulating discussions and suggestions.
I very much appreciate the initial collaboration 
with O.V. Zhirov which, hopefully, will continue before long.
I am indebted to A.A. Borovkov for elucidation of Khinchin's theorem 
and of its recent generalizations to the 'abnormal' superdiffusion.

\end{document}